\documentclass[letterpaper, 10 pt, conference]{ieeeconf}  

\IEEEoverridecommandlockouts                              

\usepackage{multirow}

\usepackage{amsthm}
\usepackage{amsmath, comment} 
\usepackage{amssymb} 
\usepackage{graphicx}
\usepackage{algorithm} 
\usepackage[noend]{algpseudocode}

\usepackage{xcolor}

\theoremstyle{definition}

\theoremstyle{remark}

\usepackage{caption}
\usepackage{subcaption}
\usepackage{url}
\usepackage{stackengine,cite}

\usepackage{cancel}

\title{\LARGE\bf Co-state Neural Network for Real-time Nonlinear Optimal Control with Input Constraints}

\author{Lihan Lian$^{1}$, Uduak Inyang-Udoh$^{2}$ \\
\thanks{$^{1}$Department of Robotics, University of Michigan, Ann Arbor, Michigan, USA. \tt\small lihanl@umich.edu}%
\thanks{$^{2}$Department of Mechanical Engineering, University of Michigan, Ann Arbor, Michigan, USA. \tt\small udiinyang@umich.edu}%
}

\begin{document}
\maketitle
\thispagestyle{empty}
\pagestyle{empty}

\begin{abstract}
In this paper, we propose a method to solve nonlinear optimal control problems (OCPs) with constrained control input in real-time using neural networks  (NNs). We introduce what we have termed co-state Neural Network (CoNN) that learns the mapping from any given state value to its corresponding optimal co-state trajectory based on the Pontryagin’s Minimum (Maximum) Principle (PMP). In essence, the CoNN parameterizes the Two-Point Boundary Value Problem (TPBVP) that results from the PMP for various initial states. The CoNN is trained using data generated from numerical solutions of TPBVPs for unconstrained OCPs to learn the mapping from a state to its corresponding optimal co-state trajectory. For better generalizability, the CoNN is also trained to respect the first-order optimality conditions (system dynamics). The control input constraints are satisfied by solving a quadratic program (QP) given the predicted optimal co-states. We demonstrate the effectiveness of our CoNN-based controller in a feedback scheme for numerical examples with both unconstrained and constrained control input. We also verify that the controller can handle unknown disturbances effectively.
\end{abstract}


\section{Introduction} \label{sec-introduction}

\subsection{Motivation}
Optimal control is a fundamental area in control theory and has wide applications across various fields including aerospace, economics and robotics \cite{teo2021applied}. In optimal control problems (OCPs), the objective is to find a feasible control input that minimizes a desired cost functional for the system of interest. There are two common strategies for solving OCPs: indirect and direct methods \cite{numerical-ocp-survey}. Indirect methods solve OCPs in continuous time based on the Pontryagin’s Minimum Principle (PMP). PMP introduces the control hamiltonian as well as differential equations and boundary conditions for both the system's state and co-state variables. This results in a Two-Point Boundary Value Problem (TPBVP) and numerical methods are often needed to obtain the optimal control policy \cite{Ascher_1995}. On the other hand, direct methods transcribe continuous-time OCPs into discrete-time Nonlinear Programming (NLP) problems via trajectory optimization techniques such as direct shooting or direct collocation. The NLP problems are then often solved by gradient-based algorithms\cite{betts_2010}.



Ideally, OCPs are solved once and for all offline. However, in the presence of disturbances and/or uncertainties, maintaining optimality requires a control scheme that utilizes real-time feedback. This demands recursively resolving a TPBVP or a NLP problem in response to feedback measurement. In practice, one of the most common strategy for implementing this is Model Predictive Control (MPC) where a discrete-time NLP problem, for a finite number of time steps or receding horizon, is solved after each feedback measurement and only the first time step input of the trajectory is executed \cite{mpc-survey}. A key challenge in MPC, and indeed for solving OCPs in real-time generally, is that for nonlinear problems there are often no guarantees of timely convergence \cite{Diehl_2005}. Moreover, solving a fresh OCP at each point in a feedback loop \textit{does not utilize the fact that patterns from previous solutions may be used to alleviate the computational burden of subsequent solutions}: hence, the need for pattern recognition / machine learning (ML) techniques in real-time OCPs. 

\subsection{Related Work}
Several studies
have leveraged neural networks (NNs) to address OCPs in various ways, deep reinforcement learning (RL) being prominent among them \cite{ABUKHALAF2005779}. In deep RL-based optimal control, NNs are used to parameterize the optimal control policy often for a discrete-time or discretized OCP. For continuous-time OCPs, optimal control policies may be parameterized using neural ODEs which are NNs that approximate differential equations \cite{neural-ode-icl, neural-ode-nips}. In either case, the NNs are trained to learn a control policy that minimizes a cost function (or functional). While several studies have shown the ability of the NNs (or neural ODEs) to learn control policies similar to those obtained using PMP or MPC \cite{ai-pontryagin-nature},  the policies learned are often only valid for the initial states specified in the OCP. In addition, most RL-based controllers cannot implicitly handle input constraints. 

Another strategy for utilizing NNs in OCPs entails imitation learning (IL) \cite{gonzalez2023neuralnetworksfastoptimisation, mpc-nn-automatica}. In this case, the NNs are trained using state and control input data pairs. The control inputs are obtained from an expert solution (often from solving an NLP or TPBVP) of the OCP. The expectation here is that the NN will learn a control law that provides a suitable control input in real-time  in a similar manner, but at a fraction of the computational cost, as the expert OCP solver. This approach has had remarkable success in linear systems with a quadratic cost where it has been shown that NNs with rectifier unit activation functions can explicitly represent associated piecewise affine optimal control laws \cite{Karg_2020, explicit-mpc-upenn, explicit-mpc-nn}. For a linear quadratic regulator (LQR) this trivially reduces to learning a gain matrix. One approach that readily extends IL-based linear control to nonlinear system is based on the State-Dependent Riccati Equation (SDRE) \cite{sdre-acc}. In the SDRE approach, the nonlinear system dynamics is converted into a state-dependent coefficient form, and then, like for an LQR problem, a suboptimal control policy is obtained by finding a NN-based parametric solution of a state-dependent Alegebraic Riccati Equation \cite{sdre-lcss,sdre-ifac}. SDRE-based control, however, does not  admit input constraints. Other IL-based approaches for nonlinear control  are typically blackbox. That is the NN controller is trained to minimize control input deviations from  the expert without insights into the underlying dynamics of the expert solver. The result is that approximation errors may become significant for states that are unencountered in training, not to mention lack of guarantees on compliance with constraints \cite{gonzalez2023neuralnetworksfastoptimisation}. 

A third strategy for using NN in OCPs entails modeling the solution to the TPBVP with a NN.  In one of the limited studies that have explored this approach, the NN is trained only to satisfy the original TPBVP differential equations \cite{pontryagin_nn-mathematics}. The trained NN in that study successfully learned optimal inputs for the OCP considered. However the trained NN only provides solution to the TPBVP for a specific initial condition. In another study \cite{Li_2019}, the NNs are used in a purely supervised approach to predict the initial co-state value in the TPBVP for an OCP without input constraints. As with IL techniques previously discussed, such blackbox approach tends to perform poorly on extrapolation, that is, when testing on data points that lie substantially outside the range of training data.

\subsection{Contribution}
In this work we use a NN to parameterize the TPBVP solution using a supervised learning approach. The NN receives the system state as input and outputs a co-state trajectory. In training, we minimize the difference between co-state trajectory predicted by the NN and that obtained from an expert solver for various system’s states. But we additionally constrain the NN co-state prediction to satisfy the PMP. As with MPC strategy, once the optimal co-state trajectory is computed from the NN, the control input is obtained using only the initial co-state value. We evaluate the trained NN as a feedback controller in a closed loop setting with system disturbance and initial states outside the range of those encountered in training. Result shows that the NN drives the system to the desired state in a similar fashion as the expert solver. To enumerate, the contribution of the paper are as follows.
\begin{itemize}
    \item We present a novel perspective of using neural networks that learn the mapping from an initial state to its corresponding optimal co-state trajectory. 
    \item Instead of handling control input constraints directly during the process of training neural network, we propose a framework for handling input constraints by extracting a NN-based co-state trajectory and simply solving a QP.
    \item We verify the effectiveness of CoNN-based controller in scenarios including unconstrained control input with unseen initial conditions, constrained control input and existence of disturbance. For a one-dimensional nonlinear system, results show that our CoNN-based controller is on par with the off-the-shelf optimization solver and robust to disturbance and extrapolation.
\end{itemize}

\subsection{Paper Structure}
This paper is organized as follows: Section II presents the background of both PMP and MPC. In Section III, we define the general optimal control problem that CoNN aims to solve. Sec. IV details the design of neural network architecture and training procedures, followed by the method of handling control input constraints and CoNN-based controller validation. Sec. V provides results and performance comparison between the CoNN-based controller and a standard optimal control solver for a numerical example. Finally, Section VI concludes the paper and outlines future directions.

\section{Background} \label{sec-background}
 Consider a continuous time OCP with the objective to minimize a cost functional $J$ described by:
\begin{subequations} \label{eq:background_ocp_formulation} 
\begin{align}
J & = \phi(x(T)) + \int_{0}^{T} L(x(t), u(t), t) \, dt, \\
\text{s.t.} \quad  & \dot{x}(t) = f(x(t), u(t), t),  \label{eq:background_ocp_formulation-state_dynamics} \\
& x(0) = x_0, \label{eq:background_ocp_formulation-initial-condition}\\
& x(t_f) = x_f, \label{eq:background_ocp_formulation-terminal-condition}\\
& u(t) \in \mathcal{U} \label{eq:background_ocp_formulation-input-constraint},
\end{align}
\end{subequations}
where \(x(t) \in \mathbb{R}^p\) represents the state vector, and \(u(t) \in \mathbb{R}^q\) is the control input. Terminal cost and running cost are represented by \(\phi(x(T))\) and \(L(x(t), u(t), t)\) respectively. The solution of the OCP also needs to satisfies the constraint imposed by the system dynamics initial condition, terminal condition and control input constraints described by  \eqref{eq:background_ocp_formulation-state_dynamics}, \eqref{eq:background_ocp_formulation-initial-condition}, \eqref{eq:background_ocp_formulation-terminal-condition}, \eqref{eq:background_ocp_formulation-input-constraint} respectively.

\subsection{Model Predictive Control (MPC)}

In practice, when real-time optimal control is needed, \eqref{eq:background_ocp_formulation} is solved using MPC, or receding horizon control. MPC uses the system model to predict the future states of the system and solve the online optimization problem in a moving horizon fashion \cite{mpc-survey}. First, the continuous system dynamics in \eqref{eq:background_ocp_formulation-state_dynamics} 
is discretized based on suitable sampling interval to yield a discrete-time system:
\begin{equation}
    x_{t+1} = f(x_t, u_t),
\end{equation}
where $x_t \in \mathbb{R}^p$ represents the state vector at time step $t \in \mathbb{Z}^+$, $u_t \in \mathcal{U} \subset \mathbb{R}^q$ is the admissible control input at time step $t$. This is used to transcribe the OCP \eqref{eq:background_ocp_formulation} into an optimization problem with finite number of decision variables, thus following the scheme of \textit{discretize then optimize} paradigm of solving OCPs \cite{nonlinear-programming-book}. The discrete analog of OCP \eqref{eq:background_ocp_formulation} may be written as:
\begin{subequations} \label{eq:mpc_formulation} 
\begin{align}
J_t^*(x_t) &= \min_{u_{t:t+P-1|t}} \ell_f(x_{t+P|t}) + \sum_{k=0}^{P-1} \ell(x_{t+k|t}, u_{t+k|t})  \\
\text{s.t.} \quad  & x_{t+k+1|t} = f(x_{t+k|t}, u_{t+k|t}), \quad k = 0, \dots, P-1  \\
& u_{t+k|t} \in \mathcal{U}, \quad k = 0, \dots, P-1 \\
& x_{t|t} = x_t,  \\
& x_{t+P|t} \in \mathcal{X}_f.
\end{align}
\end{subequations}
Here, the aim is to optimize $J_t^*(x_t)$ at each time step $t$ and input sequence $u_{t:t+P-1|t}$ are the decision variables. $x_{t+k|t}$ denotes the state vector at time step $t + k$ predicted at time step $t$, obtained by starting from the current state $x_t$. Terminal cost and stage cost are represented as $\ell_f(x_{t+P|t})$ and $\ell(x_{t+k|t}, u_{t+k|t})$ respectively, $\mathcal{X}_f$ is the terminal set and $P$ is the receding horizon. The solution of this optimization problem at time step $t$ results in an optimal input trajectory $u_{t:t+P-1|t}^* = \{u_t^{*}, \dots, u_{t+P-1|t}^{*}\}$ and only the first element of $u_{t:t+P-1|t}^{*}$ is applied to the system. This process is repeated at each time step. Solving the optimization problem at each time steps, typically involves gradient-based algorithms like sequential quadratic programming or the interior point method, \cite{fletcher_ch12}. In certain cases, the solution of the MPC may also explicitly as demonstrated in \cite{explicit-mpc-survey, explicit-constrained-lqr}.



\subsection{Pontryagin's Minimum Principle}

Pontryagin's Minimum (Maximum) Principle (PMP) is a fundamental result in optimal control theory, providing necessary conditions for optimality in systems governed by differential equations. For the problem in \eqref{eq:background_ocp_formulation}, PMP introduces the control Hamiltonian $H$, defined as:
\begin{align}
H(x(t), u(t), \lambda(t), t) &= L(x(t), u(t), t)\notag \\ 
 & + \lambda^\top(t) f(x(t), u(t), t),
\label{eq:hamiltonian}
\end{align}
where \(\lambda(t) \in \mathbb{R}^n\) is the co-state (or adjoint) vector. Constraints on state variables and co-state variables are then derived by taking the partial derivatives of $H$ as follows:
\begin{equation}
\dot{x}(t) = \frac{\partial H}{\partial \lambda},
\label{eq:state_dynamics}
\end{equation}
\begin{equation}
\dot{\lambda}(t) = -\frac{\partial H}{\partial x}.
\label{eq:costate_dynamics}
\end{equation}
Since both initial conditions and final conditions on state variables are fixed for OCP \eqref{eq:background_ocp_formulation}, a TPBVP -- which comprises the differential equations \eqref{eq:state_dynamics}, and \eqref{eq:costate_dynamics}, along with the boundary conditions \eqref{eq:background_ocp_formulation-initial-condition},  and \eqref{eq:background_ocp_formulation-terminal-condition} -- emerges . The PMP states that the optimal control \(u^*(t)\) minimize the Hamiltonian $H$:
\begin{equation}
u^*(t) = \arg\min_{u(t) \in \mathcal{U}} H(x^*(t), u(t), \lambda^*(t), t).
\label{eq:optimal_control}
\end{equation}
The significance of PMP lies in the fact that minimizing the Hamiltonian is significantly easier than the original infinite-dimensional OCP. PMP transcribes the OCP to a point-wise optimization to avoid minimizing over a function space. Rather than first discretizing the system and convert the OCP to an optimization problem with finite number of decision variables as in the case of MPC, PMP tackles OCPs in continuous time by solving the corresponding TPBVPs. This positions PMP as the foundation of indirect methods and place it in the paradigm of \textit{optimize then discretize}\cite{nonlinear-programming-book}. It is often the case that no analytical solution exists for TPBVPs, and numerical techniques such as single shooting, multiple shooting \cite{numerical-tpbvp-book}, \cite{intro-numerical-analysis}  and collocation methods \cite{collocation-method-ocp, direct-collocation-intro} are required. 


\section{Problem Statement} \label{sec-problem-statement}

For a control affine system, consider the infinite final time and fixed final state optimal control problem with quadratic running cost in continuous time as follows: 
\begin{subequations} \label{eq:ocps_formulation} 
\begin{align}
\min \quad J & = \frac{1}{2}\int_{t_0}^{t_f = \infty} \left( x^\top(t)Qx(t) + u^\top(t)Ru(t) \right) dt , \\
\text{s.t.} \quad & \dot{x}(t) = f(x(t)) + g(x(t))u(t), \\
& x(t_0) \in X\\
& x(t_f = \infty) = x_f, \\
& u(t) \in \mathcal{U},  
\end{align}
\end{subequations}
where \(x(t) \in \mathbb{R}^p\) and \(u(t) \in \mathbb{R}^q\) \footnote{Note that in the illustrative example (Sec. V), we only consider the case of $p=q=1$. But in the rest of this section, we generalize the dimensions.} are as previously defined,  $x(t_0)$ and $x_f$ specify the initial and terminal conditions respectively, $X$ is a set of possible initial conditions \footnote{Note that the problem formulation here differs from standard OCPs, where $x(t_0)$ is fixed. This is to emphasize that the OCP admits a family of solutions.}, and $f(x)$ and $g(x)$ are functions with appropriate dimensions. The running cost is defined by matrix \(Q \in \mathbb{R}^{p \times p}\) a symmetric positive semi-definite matrix, and \(R \in \mathbb{R}^{q \times q}\), a symmetric positive definite matrix. Based on the PMP, the Hamiltonian \( H \) can be expressed as:
\begin{align}
H(x(t), u(t), \lambda(t), t) &= \frac{1}{2} x^\top(t)Qx(t) + \frac{1}{2} u^\top(t)Ru(t) \notag \\
&\quad + \lambda^\top(t) \left( f(x(t)) + g(x(t))u(t)\right),
\end{align}
where $\lambda(t) \in \mathbb{R}^p$ is the co-state vector that has the same dimension as state vector $x(t)$. Following \eqref{eq:state_dynamics} and \eqref{eq:costate_dynamics}, the state equation and co-state equation are derived as follows:
\begin{equation}
    \dot{x}(t) = \frac{\partial H}{\partial\lambda} = f(x(t)) + g(x(t))u(t),
\end{equation}

\begin{align}
    \dot{\lambda}(t) = -\frac{\partial H}{\partial x} &= -Qx(t) - \left( \frac{\partial f(x(t))}{\partial x} \right)^\top \lambda(t) \notag \\
    &\quad - {\left( \frac{\partial g(x(t))}{\partial x} u(t)\right)^\top} \lambda(t).
\end{align}
After solving the optimal $x^*(t)$ and $\lambda^*(t)$ from the corresponding TPBVP, the optimal control input \(u^*(t) \) is obtained from \eqref{eq:optimal_control}. Given the quadratic cost defined by $Q$ and $R$, when there are no constraints on the control input, the optimal control law can be obtained by setting:
\begin{equation} \label{eq:unconstrained-optimal-u-partial-H-partial-u}
    \frac{\partial H}{\partial u^*} = 0 ,
\end{equation}
which yields
\begin{equation} \label{eq:unconstrained-optimal-u-expression}
    u^*(t) = -R^{-1}g^\top(x(t))\lambda^*(t).
\end{equation}
When there are constraints, the optimal control input can be obtained from:
\begin{equation} \label{eq:constrained-optimal-u}
u^*(t) = \arg\min_{u(t) \in \mathcal{U}} \left( \frac{1}{2} u^\top(t)Ru(t) + \lambda^{*\top}(t) g(x(t)) u(t) \right).
\end{equation}

In many instances, the optimal co-states cannot be solved analytically or determined a priori. Therefore, numerical methods based on the differential equations and boundary conditions must be employed to solve for the optimal co-state trajectory first. This is typically computationally expensive, and thus, hinders the use of indirect methods for real-time feedback control. In addition, numerical methods may not converge when control input constraints are present \cite{betts_2010}. Moreover, a numeric solution is only obtained for a given initial condition $x(t_0)$, but we seek for a family of solutions $\forall x(t_0) \in X$.

In this paper, we propose that the optimal co-state trajectory $\lambda(\tau)\text{ for } \tau \in [t_0, t]$ can be parameterized as a mapping from its corresponding state $x(t_0)$ using a NN. Thus, given any initial state, the predicted optimal co-state trajectory from the NN can then be used to find the optimal control input efficiently from \eqref{eq:constrained-optimal-u}. 


\section{Methodology}\label{sec-methodology}


This section outlines the approach to parameterize the mapping from a state to its corresponding optimal co-state trajectory using a NN, which we henceforth term co-state NN (CoNN). First, the CoNN architecture and its associated training procedures are explained in detail. Then, we give a description of how the CoNN-based controller is used in the presence of input constraints. Finally, a validation process is presented to assess the controller's performance in various scenarios.

\begin{figure*}[t]  
    \centering
    \includegraphics[trim=0cm 0.0cm 0.0cm -0.8cm,width=1\textwidth]{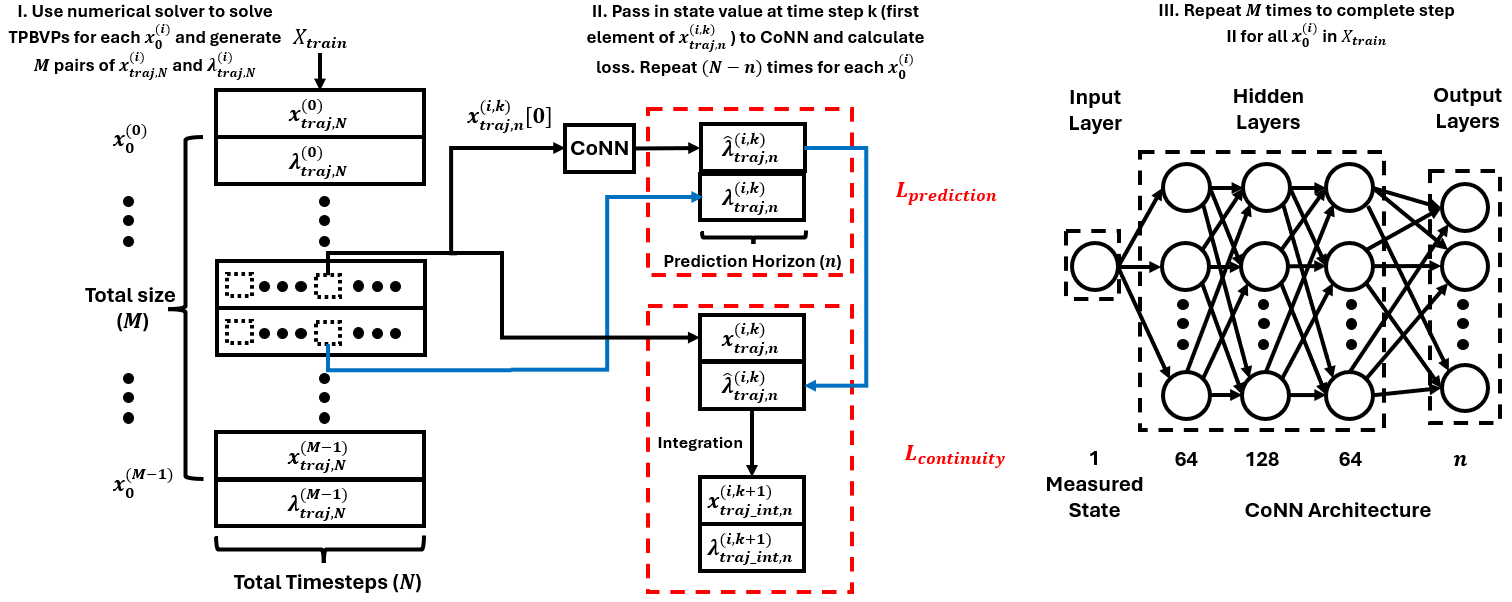}
    \caption{CoNN architecture and training procedures illustrated for $x_0^{(i)} \in \mathbb{R}, \forall i$, same as the one-dimensional system shown in the Sec. V. Note that the index [$j$] denote the $j^{th}$ entry of the trajectory starting from 0. }
    \label{fig:architecture}
\end{figure*}


\subsection{Neural Network Architecture}
For simplicity, consider the case where $x(t), \lambda(t) \in \mathbb{R}$. Furthermore, assume that our TPBVP in Sec. III is uniformly discretized with sufficiently small time intervals $\delta$ between $t_0$ and $t_f$ such that $x_k$ is the value of $x(t)$ at $t = k\delta$  for $k \in \mathbb{Z}^+$. The NN used, then, is a fully connected feedforward network, which takes $x_k$ and outputs the corresponding optimal co-state trajectory $\boldsymbol{\hat{\lambda}}_{\text{traj}, n}^{k}$ with a finite prediction horizon length of $n \in \mathbb{Z}^+$,
\begin{equation}
\boldsymbol{\hat{\lambda}}_{\text{traj}, n}^{k} = \text{CoNN}_{\theta}(x_k).
\end{equation}
where $\theta$ denotes the parameters of the NN to be optimized. Note that the bold symbol is used to emphasize that $\boldsymbol{\hat{\lambda}}_{\text{traj,n}}$ is a vector of length of $n$. The length of the co-state trajectory prediction horizon is set arbitrarily, and in a later section, it is defined as $n = 11$. For a one-dimensional system, the input layer of the CoNN consists of a single node while the output layer corresponds to the specified prediction horizon $n$ (Fig. \ref{fig:architecture}). For a $p$-dimensional system, a CoNN with output dimension $n \times p$ is needed.


\subsection{Training Procedures}
\subsubsection{Dataset generation} 

In this work, the dataset is generated by choosing evenly distributed initial state values $x_0^{(i)}$, for $i = 0, \dots, M-1$, within a prescribed range; the initial state values are stored in $X_{\text{train}}$. For each $x_0^{(i)}$, a TPBVP is solved, and both the optimal state and the co-state solution trajectories of length $N$, that is, $\boldsymbol{x}_{\text{traj, N}}^{(i)}$ and $\boldsymbol{\lambda}_{\text{traj, N}}^{(i)}$, are stored. For the infinite final time problem ($t_f = \infty$) under consideration, $N$ should be sufficiently large to ensure that the state trajectory converges.  It is important to note that the optimal solutions are obtained under the unconstrained scenario, where the trained CoNN predicts the optimal co-state trajectory in the absence of control input constraints. Once the optimal co-state trajectory is obtained, the constrained optimal  control input can then be obtained by solving a simple optimization problem (QP) as discussed in the next subsection. 
\subsubsection{Loss function} The loss function consists of two main components. The first component is the prediction loss $\mathcal{L_{\text{prediction}}}$, which is defined as the sum of mean square error (MSE) between each element in the predicted co-state trajectory and the optimal co-state trajectory obtained from a standard numerical solver. $\mathcal{L_{\text{prediction}}}$ is calculated in a \textit{receding horizon fashion}, as shown in Fig. \ref{fig:architecture}, that is, for each optimal co-state trajectory $\boldsymbol{\lambda}_{\text{traj, N}}^{(i)}$ (with length $N$) from the numerical solver, we use a sliding window approach to iteratively select the first $N-n$ shorter co-state trajectories $\boldsymbol{\lambda}_{\text{traj, n}}^{(i,k)}$ (with length $n$) for $k = 0,\dots, N-n-1$. In fact, at each time step $k$, both $\boldsymbol{x}_{\text{traj, n}}^{(i,k)}$ and $\boldsymbol{\lambda}_{\text{traj, n}}^{(i,k)}$ of length $n$ are extracted from $\boldsymbol{x}_{\text{traj, N}}^{(i)}$ and $\boldsymbol{\lambda}_{\text{traj, N}}^{(i)}$ respectively. The first element from $\boldsymbol{x}_{\text{traj, n}}^{(i,k)}$ is passed as input to CoNN and the predicted $\boldsymbol{\hat{\lambda}}_{\text{traj}, n}^{(i,k)}$ are then used to calculate MSE. $\mathcal{L}^{(i,k)}_{\text{prediction}}$ corresponds to the MSE at time step $k$ corresponds to initial condition $x_0^{(i)}$:

\begin{equation} \label{eq:L_prediction}
\mathcal{L}^{(i,k)}_{\text{prediction}} = \frac{1}{n} \sum_{j=0}^{n-1} \left( \boldsymbol{\hat{\lambda}}_{\text{traj}, n}^{(i,k)}[j] - \boldsymbol{\lambda}_{\text{traj}, n}^{(i,k)}[j] \right)^2.
\end{equation}
Thus in training, the parameters $\theta$ of the NN are updated $N-n$ times for each $x_0^{(i)}$ and $M(N-1)$ for the entire training data set $X_{\text{train}}$. 

The second component of the loss function is the continuity loss $\mathcal{L_{\text{continuity}}}$, which is calculated by using the predicted co-state trajectory $\boldsymbol{\hat{\lambda}}_{\text{traj}, n}^{(i,k)}$ from CoNN and the corresponding optimal state trajectory $\boldsymbol{x}_{\text{traj}, n}^{(i,k)}$ from the numerical solver. We integrate both the $\boldsymbol{\hat{\lambda}}_{\text{traj}, n}^{(i,k)}$  and $\boldsymbol{x}_{\text{traj}, n}^{(i,k)}$  trajectories one time step forward to produce $\boldsymbol{\lambda}_{\text{traj\_int}, n}^{(i,k+1)}$ and $\boldsymbol{x}_{\text{traj\_int}, n}^{(i,k+1)}$, both of which are of length $n$ and begin at the time step $k+1$. We calculate find the continuity loss by using the MSE defined as follows:

\begin{align} \label{eq:L_continuity}
\mathcal{L}^{(i,k)}_{\text{continuity}} &= \frac{1}{n-1} \left( \sum_{j=0}^{n-2} \left( \boldsymbol{x}_{\text{traj}, N}^{(i,k)}[j+1] - \boldsymbol{x}_{\text{traj\_int}, n}^{(i,k+1)}[j] \right)^2 \right. \notag \\
&\quad + \left. \sum_{j=0}^{n-2} \left( \boldsymbol{\hat{\lambda}}_{\text{traj}, n}^{(i,k)}[j+1] -  \boldsymbol{\lambda}_{\text{traj\_int}, n}^{(i,k+1)}[j] \right)^2 \right).
\end{align}

For a NN that adheres to the state and co-state dynamics, the predicted co-state trajectory should produce a  small continuity loss $\mathcal{L}^{(i,k)}_{\text{continuity}}$ for all $i, k$. The goal of the continuity loss is, essentially, to infuse the NN with information about system dynamics. The full training procedure is outlined in Algorithm \ref{alg:alg1} and described in Fig. \ref{fig:architecture}.

\begin{algorithm}
\caption{CoNN Training Procedures}\label{alg:alg1}
\begin{algorithmic}[1] 
\Procedure {Training Set Generation}{}
    \State \textbf{Define} set of initial conditions $X_{\text{train}}$, prediction horizon $n$, total time steps $N \geq n$
    \State \textbf{For} $x_0^{(i)} \in X_{\text{train}}$ \textbf{do}
        \State \quad Solve the TPBVP results from Eq. \eqref{eq:ocps_formulation} and PMP by numerical solvers
        \State \quad Store optimal state solution trajectory $\boldsymbol{x}_{\text{traj, N}}^{(i)}$
        
        \State \quad Store optimal co-state solution trajectory $\boldsymbol{\lambda}_{\text{traj, N}}^{(i)}$
    \State \textbf{End for}
    
\EndProcedure
\Procedure {CoNN Training}{}
    \State \textbf{Define} total training epochs $N_{\text{epoch}}$, learning rate $\alpha$ and initialize CoNN parameters $\theta$
    \State \textbf{For} $e$ in $range(N_{\text{epoch}})$ \textbf{do}
        \State \quad \textbf{For} $x_0^{(i)} \in X_{\text{train}}$ \textbf{do}
            \State \quad \quad \textbf{For} $k$ in $range(N-n)$ \textbf{do}
                \State \quad \quad \quad Get $\boldsymbol{\lambda}_{\text{traj, n}}^{(i,k)}$ and $\boldsymbol{x}_{\text{traj, n}}^{(i,k)}$
                \State \quad \quad \quad $\hat{\boldsymbol{\lambda}}_{\text{traj}, n}^{(i,k)} = \text{CoNN}_{\theta}(\boldsymbol{x}_{\text{traj, n}}^{(i,k)}[0])$
                \State \quad \quad \quad Calculate $\mathcal{L}^{(i,k)}_{\text{prediction}} $ and $\mathcal{L}^{(i,k)}_{\text{continuity}}$ based on Eq. \eqref{eq:L_prediction} and \eqref{eq:L_continuity}
                \State \quad \quad \quad \textbf{Update} CoNN parameters $\theta$
            \State \quad \quad \textbf{End for}
        \State \quad \textbf{End for}
    \State \textbf{End for}
\EndProcedure
\end{algorithmic}
\end{algorithm}

\subsection{Control Input Constraints Handling}

Given that the running cost function of the OCP \eqref{eq:ocps_formulation} is quadratic with respect to the state $x$ and control input $u$, in the absence of input constraints, the optimal control input can be obtained based on \eqref{eq:unconstrained-optimal-u-expression}. In the case of constrained inputs, the optimal control input is determined by minimizing the Hamiltonian $H$ over the admissible set $\mathcal{U}$ as indicated in \eqref{eq:constrained-optimal-u}. In a real-time feedback control loop, CoNN takes the measured state value as the input and predicts the optimal co-state trajectory $\hat{\boldsymbol{\lambda}}_{\text{traj}, n}^{k}$ with a length of $n$. To obtain the discrete-time (sub)optimal control input $u^*_k$ at time step $k$, only the first element of the predicted trajectory  $\hat{\boldsymbol{\lambda}}_{\text{traj}, n}^{k}$, denoted as $\hat{\boldsymbol{\lambda}}_{\text{traj}, n}^{k}[0]$, is needed to perform a QP as shown in \eqref{eq:easier_qp}.


\begin{subequations} \label{eq:easier_qp} 
\begin{align}
    u^*_k &= \arg\min \left( \frac{1}{2} u_k^\top R u_k + (\hat{\boldsymbol{\lambda}}_{\text{traj}, n}^{k}[0])^\top g(x(t)) u_k \right), \\
    \text{s.t.} \quad & \hat{\boldsymbol{\lambda}}_{\text{traj}, n}^{k} = \text{CoNN}_{\theta}(x_k),  \\
    & u_k \in \mathcal{U}.
\end{align}
\end{subequations}
\subsection{CoNN-based Controller Validation}

\begin{figure}[h!]
    \centering    \includegraphics[trim=0cm 0.0cm 0.0cm -0.8cm,width=0.5\textwidth]{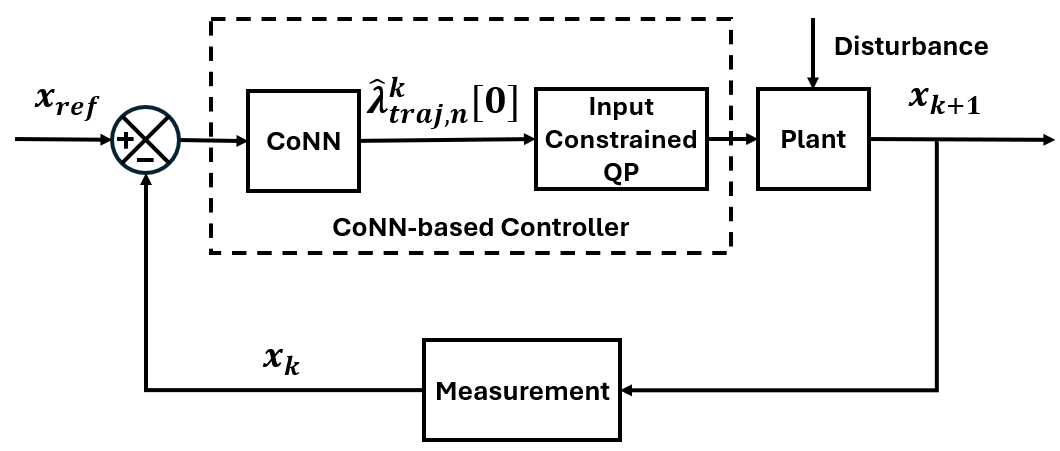}
    \caption{CoNN-based controller block diagram for validation.}
    \label{fig:validation}
\end{figure}
To evaluate the performance of the trained CoNN, validation is conducted in a real-time control loop, as illustrated in Fig. \ref{fig:validation}. Once the CoNN receives the state value $x_k$ as input at time step $k$, it predicts the optimal co-state trajectory with a length of $n$, as defined previously. The first element of the predicted co-state trajectory is then extracted and used to solve an easier QP based on \eqref{eq:easier_qp}. This process accounts for control input constraints and generates the optimal control input $u_k^*$ for the current time step $k$. The computed control input is applied to the system, and then an integrator is used to advance the system by one time step. As shown in Fig. \ref{fig:validation}, we also consider that the system may be subjected to unknown disturbance.

{\textit{Remark :}} (1) The mapping that the CoNN learns is based on the indirect method or \textit{optimize then discretize} paradigm, but the CoNN-based controller operates in a similar fashion as MPC. Thus, our CoNN-based controller can be viewed as an indirect-method-based MPC with reduced computational burden.\\
(2) Note that, in our approach, irrespective of the prediction horizon $n$,  the CoNN is trained (based on PMP) to produce trajectories and control inputs that are optimal with respect to the original OCP at each time step. This is more or less at variance with the direct MPC approach, which can only produce control inputs that are optimal only for the portion of the OCP (depending on receding horizon $P$) considered at each time step. 


\section{Example}\label{sec-example}
Consider the following one-dimensional nonlinear optimal control problem in continuous time that has quadratic running cost:
\begin{subequations} \label{eq:example_problem} 
\begin{align}
\min_{u} \quad J  & = \frac{1}{2} \int_{0}^{\infty} \left( x^2 + u^2 \right) \, dt ,\\
\text{s.t.} \quad & \dot{x} = -x^2 + x + u,\\
& u_{\text{min}} \leq u \leq u_{\text{max}}, \\
& x_0 \in \mathbb{R}, \\
& x_{\infty} = 0,
\end{align}
\end{subequations}
where $X_{\text{train}}$, is chosen to be $[-5.0, 5.0]$ for the range of initial conditions that produce TPBVPs. A total of 101 data points were sampled uniformly within the range of $X_{\text{train}}$ for a numerical solver to generate corresponding pairs of optimal state and co-state trajectories. Based on observation, the system state trajectories converge to $0$ within $10$ second for all initial conditions in the unconstrained scenario. We then took $10$ second as the final time $t_f$ of integration and set uniform time increment $\delta = 0.05$ second for the numerical solver. Thus, both state and co-state trajectories have the length of total time steps $N = 201$. Given an initial condition, $solve\_bvp$ function from $scipy$ is used as the numerical solver to subsequently produce a corresponding pair of optimal state and co-state trajectories. The prediction horizon $n$ for co-state trajectory is set as 11. Thus, the CoNN for this case takes one state value and output a vector that has a length of $11$.

Based on PMP, the control Hamiltonian $H$ can be expressed as follows:
\begin{equation}
H = \frac{1}{2} x^2 + \frac{1}{2} u^2 + \lambda \left( -x^2 + x + u \right).
\end{equation}
We can then derive the differential equation that optimal state and costate should follow:
\begin{equation}
\dot{x}^* = \frac{\partial H}{\partial \lambda} = -x^{*2} + x^* + u^*,
\end{equation}
\begin{equation}
\dot{\lambda}^* = - \frac{\partial H}{\partial x} = - \left( x^* - 2 \lambda^* x^* + \lambda^* \right).
\end{equation}
When there is no constraint on control input, the optimal control policy can be determined based on \eqref{eq:unconstrained-optimal-u-partial-H-partial-u} and \eqref{eq:unconstrained-optimal-u-expression},
\begin{equation}
\frac{\partial H}{\partial u^*} = u^* + \lambda^* = 0 ,
\end{equation}
\begin{equation}
u^* = -\lambda^* .
\end{equation}
When there are control input constraints, the optimal control input should minimize the control Hamiltonian $H$. For this problem formulation, $H$ is a quadratic function of $u$, thus, the optimal control policy can be obtained by simply saturate the unconstrained optimal control input:
\begin{equation} \label{eq:constrained_optimal_u_expression}
u^* = \text{Sat}_{u_{\text{min}}}^{u_{\text{max}}}(-\lambda^*) .
\end{equation}

A CoNN is then trained by following the procedures illustrated in Algorithm \ref{alg:alg1}. After training, we validate the performance of the trained CoNN by testing it in a realtime control loop shown in Fig. \ref{fig:validation}, where the CoNN takes the state value at every time step and only the first value from the predicted co-state trajectory is used for obtaining the optimal control input for that particular time step. As will be shown in the following subsections, our trained CoNN-based controller is on par with the off-the-shell numerical and optimization solver in both unconstrained and constrained control input cases and is able to make the state trajectory converge to the desired final state $x_{\text{target}} = 0$ even with the existence of disturbance.

\subsection{Unconstrained Control Input with Unseen $x_0$}

\begin{figure}[h!]
\begin{subfigure}[b]{\columnwidth}
   \begin{center}
   \includegraphics[trim=0cm 0.0cm 0.0cm 0.0cm,width=1\textwidth]{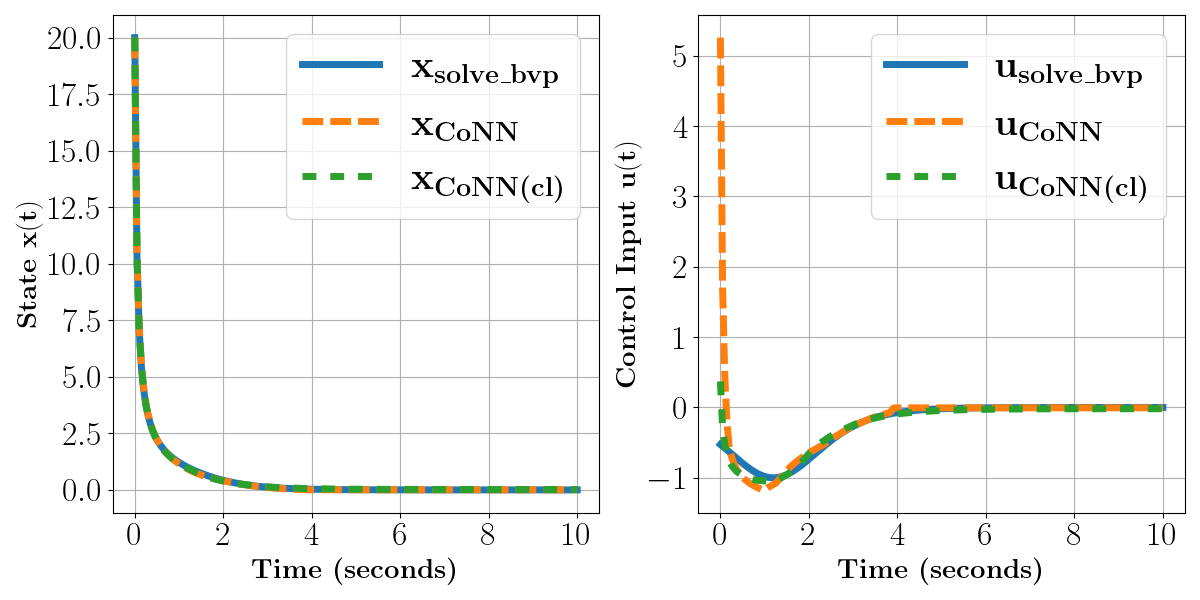}
   \caption{Initial condtion $x_0 = 20$.}
   \label{fig:trajectory_unconstrained1}
   \end{center}
\end{subfigure}
\begin{subfigure}[b]{\columnwidth}
   \begin{center}
      \includegraphics[trim=0cm 0cm 0cm 0cm, width=1\textwidth]{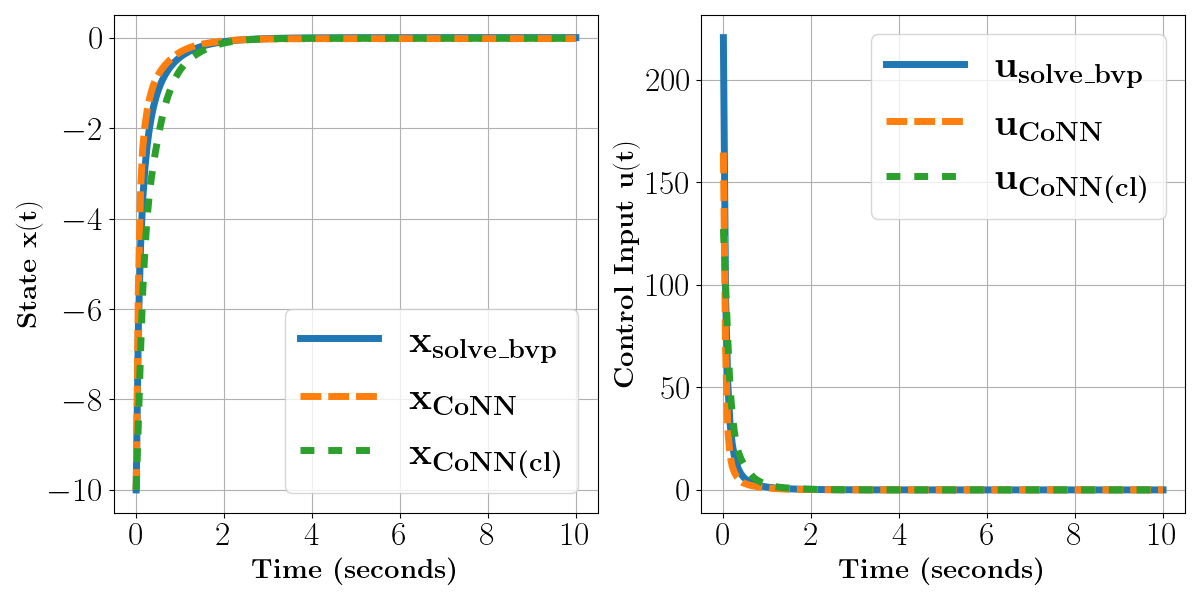}
\caption{Initial condition $x_0 = -10$.}
   \label{fig:trajectory_unconstrained2}
   \end{center}
\end{subfigure}
\caption[8pt]{CoNN-based controller with unseen initial conditions.}\label{fig:trajectory_unconstrained}
\end{figure}

To evaluate the extrapolation performance of CoNN, we test it in the unconstrained scenario with unseen initial states $(x_0 \notin X_{\text{train}})$. As shown in Fig. \ref{fig:trajectory_unconstrained}, both CoNN trained with ($x_{\text{CoNN}(cl)}$) and without continuity loss ($x_{\text{CoNN}}$) converges to $x_{\text{target}} = 0$, for unseen $x_0 = 20$ and $x_0 = -10$. The model trained with continuity loss tends to be more conservative and the model trained with out continuity loss tends to be more aggressive. Compared with the optimal control input from the numerical solver ($u_{\text{solve\_bvp}}$), the model trained with continuity loss provides a significantly better approximation in case where $x_0 > 5$ , but has slightly worse approximation for $x_0 < -5$ due to its conservativeness. This is illustrated in Fig. \ref{fig:trajectory_unconstrained} for $x_0 = 20$ and $x_0 = -10$. In this problem, choosing negative initial condition result in significant demand on control input in unconstrained scenario, as depicted in the comparison between Fig. \ref{fig:trajectory_unconstrained1} and Fig. \ref{fig:trajectory_unconstrained2}. Hence it is natural to apply input constraints in real application.

\subsection{Constrained Control Input}
\begin{figure}
    \centering
    \includegraphics[width=0.5\textwidth]{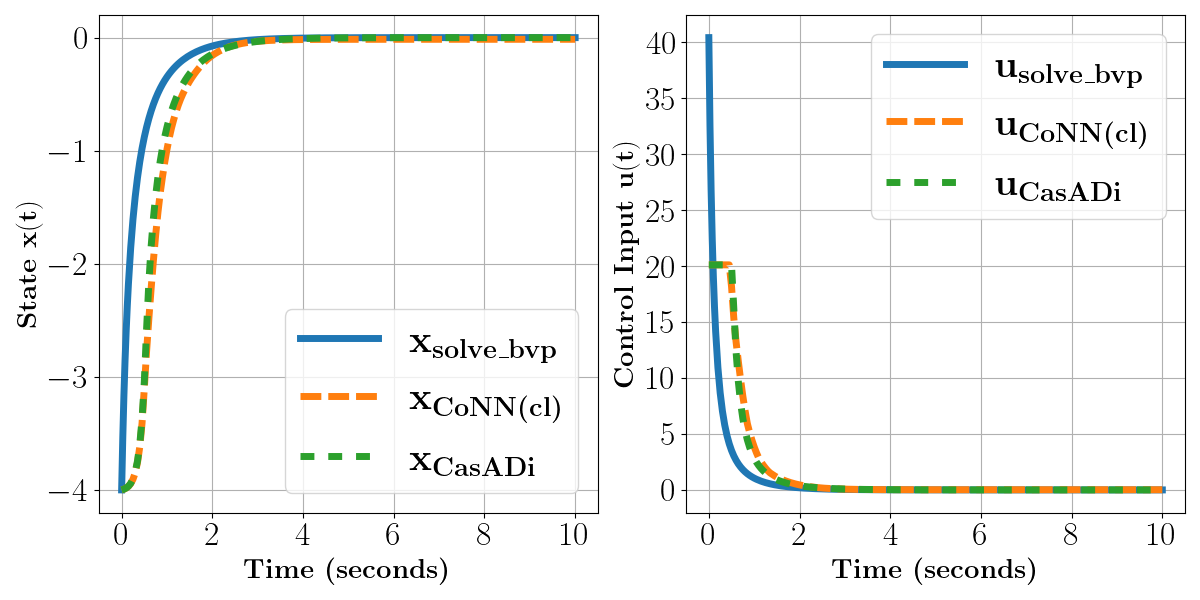}
    \caption{CoNN-based controller with control input constraints}
    \label{fig:trajectory_constrained}
\end{figure}

To validate the performance of CoNN in the constrained control input scenario, we chose $x_0 = -4$ as the initial condition. Given the system dynamics $\dot{x} = -x^2 + x + u$ and the target state $x = 0$, when the initial condition is $x(0) = -4$, the system equation becomes $\dot{x} = -20 + u$. For the state $x$ to move towards zero, $\dot{x}$ must be positive. Thus, the control input $u$ must satisfy $u > 20$ to ensure that $\dot{x} > 0$. If the maximum allowable control input is less than or equal to 20, the state $x$ will not be able to achieve the target state, making the control objective infeasible under such constraints. Subsequently, $u_{\text{max}} = 20.1$ and $u_{\text{min}} = -20.1$ are chosen to be the input limit. Since $H$ is quadratic with respect to $u$ and $u^* = -\lambda^*$, simple saturation function can be applied to achieve optimal input as expressed in \eqref{eq:constrained_optimal_u_expression}.

For comparison purpose, a direct method is also employed to solve the problem. Direct collocation is one of the most popular direct methods for trajectory optimization \cite{direct-collocation-intro}. It discretizes the state and control input trajectories to make them both as decision variables, and then transcribes the problem into a NLP problem. Trapezoidal approximation is used for imposing the dynamics constraints, with CasADi serving as the optimization solver. As depicted in Fig. \ref{fig:trajectory_constrained}, the solution from our CoNN is consistent with the solution obtained through the direct method. 

\subsection{Validation with Disturbance}
\begin{figure}
    \centering
    \includegraphics[trim=0cm 0.0cm 0.0cm -0.6cm, width=0.5\textwidth]{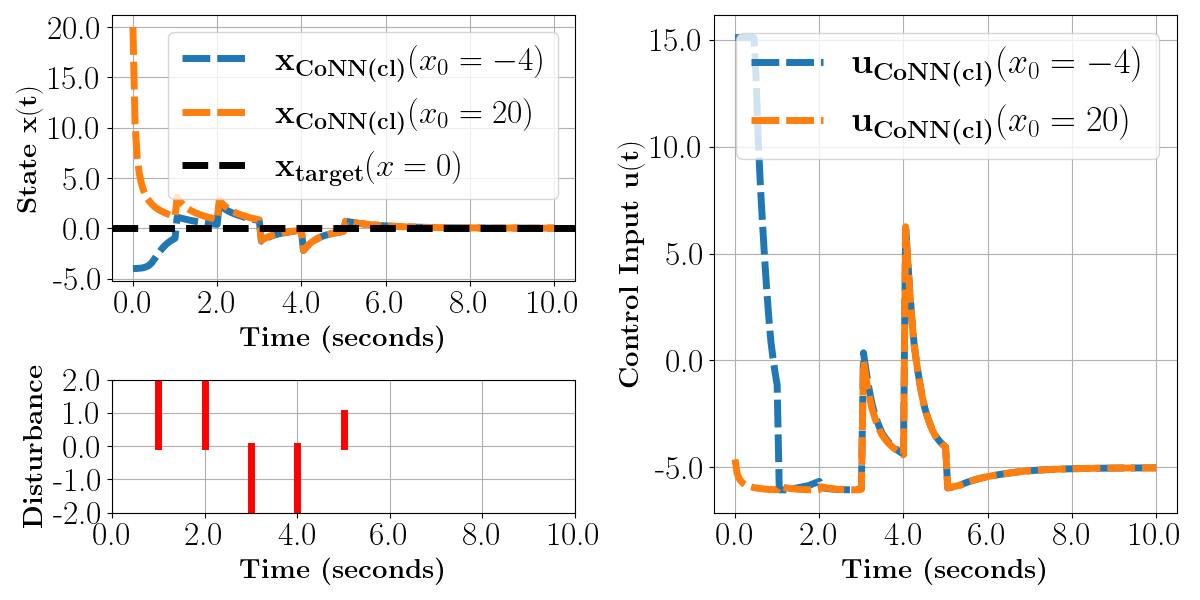}
    \caption{CoNN-based controller with disturbance}
    \label{fig:trajectory_constrained_with_disturbance}
\end{figure}

Disturbances are also injected during validation to test the robustness of our CoNN-based controller. The system is perturbed by a positive disturbance with magnitude of 2 at $t = 1, 2$ second, a negative disturbance with magnitude of 2 at $t = 3, 4$ second and a positive disturbance with magnitude of 1 at $t = 5$ second. For both scenarios of constrained control input with initial condition $x_0 = -4.0$ (seen) and unconstrained control input with initial condition  $x_0 = 20.0$ (unseen), the state trajectory successfully converges at $t = 10$ seconds, as depicted in Fig. \ref{fig:trajectory_constrained_with_disturbance}.
\section{Conclusion}\label{sec-conclusion}
In this work, we present an approach for solving a nonlinear constrained optimal control problem from a novel perspective, leveraging the neural network to learn the mapping from a state to its corresponding optimal co-state trajectory. This not only enhances the handling of control input constraints more efficiently, but it can also be employed in a feedback scheme to make NN-based controller more robust to unseen initial conditions and disturbance. Future works will focus on extending the neural network architecture for higher-dimensional systems and incorporating state constraints into the training design.





\bibliographystyle{IEEEtran}
\bibliography{ref}

\end{document}